\documentclass[conference]{IEEEtran}
\IEEEoverridecommandlockouts
\usepackage{amsmath,amssymb,amsfonts,mathtools}
\usepackage{bm}
\usepackage{graphicx}
\usepackage{booktabs}
\usepackage{multirow}
\usepackage{algorithm}
\usepackage{algpseudocode}
\usepackage{siunitx}
\usepackage{xcolor}
\usepackage{url}
\usepackage[normalem]{ulem}
\usepackage{cite}
\usepackage{subfigure}
\usepackage{array}
\usepackage{dsfont}

\def\BibTeX{{\rm B\kern-.05em{\sc i\kern-.025em b}\kern-.08em
    T\kern-.1667em\lower.7ex\hbox{E}\kern-.125emX}}
\usepackage{balance}

\begin{document}

\title{Diffusion-Aided Bandwidth-Efficient Semantic Communication with Adaptive Requests
}

\author{
    \IEEEauthorblockN{
    Xuesong~Wang\IEEEauthorrefmark{1}\IEEEauthorrefmark{4}, Xinyan~Xie\IEEEauthorrefmark{2}, Mo~Li\IEEEauthorrefmark{1}, and Zhaoqian~Liu\IEEEauthorrefmark{1}\\
    {wangxuesong@cuhk.edu.cn}\IEEEauthorrefmark{4}
    }
    \IEEEauthorblockA{\IEEEauthorrefmark{1}The Chinese University of Hong Kong, Shenzhen, Shenzhen 518172, China.}
    \IEEEauthorblockA{\IEEEauthorrefmark{2}College of Smart Materials and Future Energy, Fudan University, Shanghai 200433, China.}
}

\markboth{}
{Diffusion-Aided Bandwidth-Efficient Semantic Communication with Adaptive Requests}

\maketitle


\begin{abstract}

Semantic communication focuses on conveying the task-relevant meaning rather than exact bitwise recovery. For image transmission with a generative receiver, relying only on text descriptions can be insufficient to preserve instance-specific visual evidence, whereas sending dense latent representations can incur substantial overhead. This paper presents a receiver-driven closed-loop scheme that transmits a short caption together with an initial sparse subset of latent blocks, and then uses feedback to request additional blocks only when needed. At each round, the receiver reconstructs the image via latent diffusion inpainting and applies a semantic consistency check between a caption generated from the reconstruction and the received caption, using a lightweight language similarity score such as ROUGE-L. The receiver stops early once a target consistency level is met, and otherwise requests a small number of additional latent blocks to refine the reconstruction. Experiments on Flickr30k over AWGN channels demonstrate a controllable rate-quality tradeoff. Adaptive feedback achieves the strongest semantic alignment and the lowest failure rate, outperforming budget-matched one-shot transmission while typically using fewer latent blocks than always-on retransmission.

\end{abstract}

\begin{IEEEkeywords}
Semantic communication, diffusion model, image inpainting, automatic repeat request.
\end{IEEEkeywords}


\section{Introduction}

Traditional communication systems aim to reliably transmit bits under channel constraints. In many scenarios, however, what ultimately matters is the utility of the received content instead of exact bitwise accuracy \cite{sagduyu2023task}. This has motivated semantic communication, in which a transmitter conveys a representation tailored to preserve task-relevant meaning, with system performance evaluated in the task space \cite{liu2024survey, wang2025diffusion}. Such systems can be implemented with end-to-end neural network approaches that train encoders and decoders simultaneously, with a task model optionally integrated \cite{liu2024explainable, grassucci2023generative}. For image transmission, sending pixels demands heavy transmission bandwidth while compact semantics enable coherent reconstructions even at low bitrate \cite{tung2025multi}.

Generative models further strengthen this paradigm by providing a powerful prior at the receiver. Latent diffusion models, for instance, can synthesize high-quality images from captions and/or latent inputs, enabling the transmission of latents instead of pixels to reduce bandwidth \cite{blattmann2023align, rombach2022high}. However, replacing pixels with full latents alone does not eliminate the rate bottleneck, since latents may still occupy a substantial budget, and text-only conditioning can yield reconstructions that drift from the source content \cite{cicchetti2024language, wang2025trustworthy}. This motivates a variable-length strategy that transmits an initial sparse set of latent blocks together with a short caption, and uses receiver feedback to request additional blocks only when the current reconstruction is not semantically sufficient. A pretrained caption-conditioned latent diffusion inpainting model can leverage such sparse evidence by filling missing regions and maintaining global coherence \cite{xie2023smartbrush}, while language guidance steers generation toward the desired content with minimal overhead \cite{manukyan2023hd}.

This strategy naturally requires a feedback mechanism that decides when to request more blocks and when to stop. In wireless settings, a fixed one-shot budget is problematic for two reasons. First, the receiver does not know the ground-truth image and therefore cannot determine in advance how much visual evidence is sufficient for the intended semantics. Second, image difficulty varies substantially, since some images can be reconstructed with few blocks whereas others require more evidence, so a fixed budget either wastes blocks or leaves the reconstruction semantically inconsistent. 
Classical hybrid automatic repeat request (HARQ) adapts redundancy via feedback to achieve bit reliability. However, semantic image transmission with a generative receiver lacks access to the source image at the receiver, so a bit-level correctness check is not available as a semantic stopping signal.
Prior studies incorporate HARQ into semantic transmission via incremental knowledge-based designs \cite{zhou2022adaptive, liu2024adaptive} or via task/feature-driven error detection and triggering mechanisms \cite{fei2025topology, li2025semantic, sheng2025semantic}. However, they do not explicitly formulate a receiver-side semantic stopping rule for generative reconstruction when the source image is unavailable, based on a thresholded semantic score computed from the receiver’s reconstruction and the received intent that directly governs both early stopping and retransmission triggering.

Motivated by these gaps, we develop a receiver-side semantic control rule for variable-length transmission with early stopping under a tunable semantic target, and evaluate it under a matched average latent-block budget to isolate the benefit of semantic control. Our contributions are listed as follows:

\begin{figure*}[!t]
    \centering
    \includegraphics[width=\linewidth]{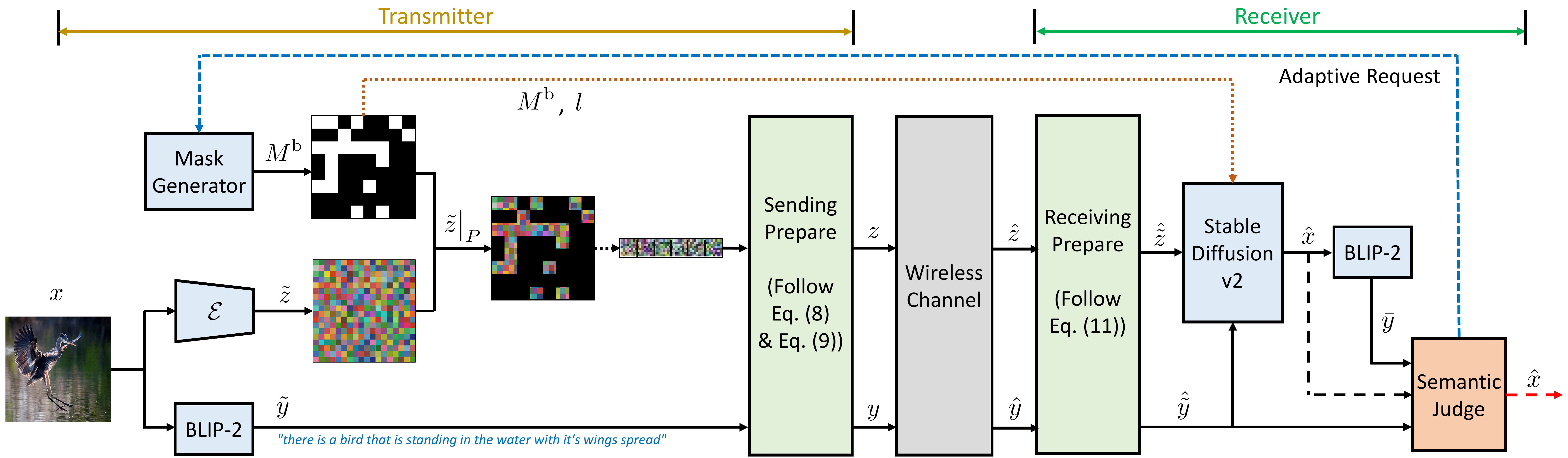}
    \caption{End-to-end workflow of the proposed receiver-driven semantic retransmission system.}
    \label{fig:whole-system}
\end{figure*}

\begin{itemize}
    \item We propose a receiver-side semantic trigger and early-stopping rule that enables variable-length transmission without access to the source image at the receiver.
    \item We couple incremental latent-block requests with caption-conditioned latent diffusion inpainting for progressive reconstruction refinement.
    \item We improve robustness under noisy links by selecting between guided and unguided reconstructions using the same semantic-consistency score.
\end{itemize}

\section{System Model}


We consider point-to-point image transmission over a wireless channel (Fig.~\ref{fig:whole-system}). The transmitter $\mathcal{T}$ sends a caption and sparse latent blocks, while the receiver $\mathcal{R}$ reconstructs by diffusion inpainting and requests additional blocks via feedback. Our goal is to reduce the expected latent payload while limiting semantic inconsistency under a receiver-side criterion.

\subsection{Transmitter-side Design}

Let $x \in \mathbb{R}^{C\times H\times W}$ denote the input image from a given dataset. A latent encoder $\mathcal{E}$ maps $x$ to a latent embedding
$\tilde{z} = \mathcal{E}(x)\in \mathbb{R}^{C_L \times H_L \times W_L}$.
The encoded latent is uniformly partitioned into square blocks. Given a block side length $l\in\mathbb{N}$ such that $l$ divides both $H_L$ and $W_L$, we can get
\begin{equation}
N_H = \frac{H_L}{l},\qquad N_W = \frac{W_L}{l},\qquad N = N_H N_W ,
\end{equation}
and index the $N$ blocks by $(i,j)$ with $i\in\{1,\dots,N_H\}$ and $j\in\{1,\dots,N_W\}$.
Each block contains all channels, that is, block $(i,j)$ has shape $C_L\times l\times l$. For notational simplicity, we omit the channel index $C_L$ in what follows. All operations apply identically to every channel.

Define a binary block-level mask
\begin{equation}
M^{\mathrm{b}}\in\{0,1\}^{N_H\times N_W},
\end{equation}
where $M^{\mathrm{b}}_{ij}=0$ marks block $(i,j)$ for transmission, and $M^{\mathrm{b}}_{ij}=1$ marks it for omission with receiver-side inpainting.
Lifting the block mask to the latent spatial grid yields
\begin{equation}
\label{eq:M_from_Mb}
M = M^{\mathrm{b}}\otimes\mathbf{1}_{l\times l}\ \in\{0,1\}^{H_L\times W_L},
\end{equation}
where $\otimes$ denotes the Kronecker product and $\mathbf{1}_{l\times l}$ is the $l\times l$ all-ones matrix. Hence $M(u,v)=0$ on transmitted regions of $\tilde{z}$ and $M(u,v)=1$ on withheld regions of $\tilde{z}$, where $u\in\{1,\dots,H_L\}$ and $v\in\{1,\dots,W_L\}$.
A random block mask is drawn under a prescribed withheld ratio
\begin{equation}
d \triangleq \frac{1}{N}\sum_{i=1}^{N_H}\sum_{j=1}^{N_W}M^{\mathrm{b}}_{ij}\ \in[0,1],
\end{equation}
namely the fraction of blocks that are not transmitted, and the block-level sending rate is $\kappa=1-d$.

Let
\begin{equation}
\label{eq:PQ}
\begin{aligned}
P &= \{(i,j):\ M^{\mathrm{b}}_{ij}=0\},  \quad
Q &= \{(i,j):\ M^{\mathrm{b}}_{ij}=1\},
\end{aligned}
\end{equation}
be the sets of transmitted and withheld (dropped) blocks, respectively.
The $M^{\mathrm{b}}_{ij}=0$ blocks are serialized and modulated into baseband symbols, given by
\begin{equation}
z = \mathcal{M}\bigl(\mathcal{C}\bigl(\mathcal{S}(\tilde{z}\big|_{P})\bigr)\bigr),
\end{equation}
where $\mathcal{S}(\cdot)$ serializes the selected latent blocks into a bitstream
using IEEE 754-2008 standard (64 bits per coefficient), $\mathcal{C}$ is a rate-$1/2$ convolutional code, $\mathcal{M}(\cdot)$ maps bits to symbols using 16-QAM, and $\tilde{z}\big|_{P}$ denotes the collection of blocks indexed by $P$ to be transmitted. The text stream is short but important because it conditions generation and provides the semantic reference for receiver-side control. We therefore transmit the caption using a standard digital chain with lightweight channel coding to improve caption reliability while incurring only a small overhead.

Let $\tilde{y}=\mathcal{F}(x)$ denote the caption produced by an image-to-text converter. The caption is converted to bits by ASCII mapping $\mathcal{A}(\cdot)$, encoded by $\mathcal{C}(\cdot)$, and modulated by $\mathcal{M}(\cdot)$ as
\begin{equation}
y = \mathcal{M}\bigl(\mathcal{C}\bigl(\mathcal{A}(\tilde{y})\bigr)\bigr).
\end{equation}
The two streams are asymmetric: the caption is short but affects semantic control while the latent blocks carry most visual evidence. In experiments, we vary the signal-to-noise ratio (SNR) to emulate different link qualities under the same receiver-side reconstruction procedure, while leaving noise-aware latent refinement \cite{wu2024cddm} to future work.

Finally, both $z$ and $y$ are power-normalized and are sent into the same additive white Gaussian noise (AWGN) channel, given by
$\hat{z}=z + n_z$ and $\hat{y}=y + n_y$, where $n_z,n_y \sim \mathcal{CN}\bigl(0,\sigma^2\mathbf{I}\bigr)$.


\subsection{Receiver-side Design}


The receiver $\mathcal{R}$ uses the current payload $(\hat{z},\hat{y})$ to reconstruct an image estimate $\hat{x}$ under the current withholding mask $M$.
We assume the block mask $M^{\mathrm{b}}$ and block size $l$ are conveyed as reliable control metadata (hence $M$ and $d$ are known), and $\mathcal{R}$ has an accurate SNR estimate.

Several preparatory operations are applied to $\hat{z}$ and $\hat{y}$ before image reconstruction, denoted as
\begin{equation}
\label{eq:received_zy}
\begin{aligned}
\hat{\tilde{z}} &= \oplus\left\{\mathcal{S}^{-1}\Bigl(\mathcal{C}^{-1}\bigl(\mathcal{M}^{-1}(\hat{z})\bigr)\Bigr)\Big|_{P}; M \right\}, \\
\hat{\tilde{y}} &= \mathcal{A}^{-1}\Bigl(\mathcal{C}^{-1}\bigl(\mathcal{M}^{-1}(\hat{y})\bigr)\Bigr),
\end{aligned}
\end{equation}
where $\mathcal{M}^{-1}$, $\mathcal{S}^{-1}$, $\mathcal{C}^{-1}$ and $\mathcal{A}^{-1}$ are the inverse operations of $\mathcal{M}$, $\mathcal{S}$, $\mathcal{C}$ and $\mathcal{A}$, respectively.
The operator $\oplus\{\cdot;M\}$ embeds the recovered blocks into a full latent of size $C_L\times H_L\times W_L$ according to the block mask $M$, that is
\begin{equation}
\bigl[\oplus\{U;M\}\bigr](c,u,v)=
\left\{
\begin{array}{@{}l@{\;}l@{}}
U(c,u,v), & \text{if } M(u,v)=0,\\[-1pt]
0,        & \text{if } M(u,v)=1,
\end{array}
\right.
\end{equation}
where $c\in\{1,\dots,C_L\}$.

Next, we reconstruct via caption-conditioned latent diffusion inpainting using a DDIM sampler $\Gamma$ \cite{song2020denoising}, which clamps known sites and updates only masked sites.
For notational simplicity, let $\tilde{z}^{0}\coloneqq\hat{\tilde{z}}$, and let $\tilde{z}^{n}$ denote the latent at iteration $n$. With a text-conditioned denoiser $\epsilon_{\theta}(\cdot;\hat{\tilde{y}})$ and a noise schedule $\bar{\alpha}_{n}\in(0,1]$, the DDIM proposal (without the mask constraint) is
\begin{equation}
\begin{aligned}
\tilde{z}_{\mathrm{f}}^{n-1}
&=
\sqrt{\bar{\alpha}_{n-1}}
\frac{\tilde{z}^{n}-\sqrt{1-\bar{\alpha}_{n}}\epsilon_{\theta}\bigl(\tilde{z}^{n};\hat{\tilde{y}}\bigr)}{\sqrt{\bar{\alpha}_{n}}}
\\
&\quad+\sqrt{1-\bar{\alpha}_{n-1}}
\epsilon_{\theta}\bigl(\tilde{z}^{n};\hat{\tilde{y}}\bigr).
\end{aligned}
\end{equation}
The inpainting constraint then projects this proposal by keeping transmitted blocks fixed and updating only withheld sites:
\begin{equation}
\tilde{z}^{\,n-1}
=
(\mathbf{1}-M)\odot \tilde{z}^{\,n-1}_{\mathrm{r}}
\;+\;
M\odot \tilde{z}^{\,n-1}_{\mathrm{f}},
\end{equation}
where $\tilde{z}^{\,n-1}_{\mathrm{r}}$ is the decoded latent propagated to step $n\!-\!1$ by the DDIM forward map, given by
\begin{equation}
\tilde z_{\mathrm r}^{\,n-1}
\triangleq
\sqrt{\bar{\alpha}_{n-1}}\,\hat{\tilde z}
+
\sqrt{1-\bar{\alpha}_{n-1}}\,\epsilon,
\end{equation}
with $\epsilon\sim\mathcal{N}(0,\mathbf{I})$ sampled once per reconstruction (fixed by the random seed) and reused across DDIM steps.
Here, $\odot$ means element-wise multiplication with $M$ broadcast along the channel dimension, and $\mathbf{1}$ is an all-ones matrix matching $M$'s shape.

Two parameters regulate the decoder’s trade-off between caption conditioning and reliance on received blocks: the guidance scale $w$ and the inpainting strength $s_t$. We fix $w$ in our experiments. At round $t$, define $d_t \triangleq |Q_t|/N$ and set $s_t \leftarrow s(\mathrm{SNR}, d_t)$ by a simple heuristic, where $s_t$ decreases with SNR and increases with $d_t$. The sampler $\Gamma$ then runs $S_t \triangleq \lceil s_t T \rceil$ DDIM steps (with $T$ the maximum step budget) and the VAE decoder $\mathcal{D}$ maps the resulting latent to the reconstructed image $\hat{x}$.

\addtolength{\topmargin}{0.06in}
\begin{algorithm}[t]
\caption{Receiver-side semantic-driven adaptive retransmission.}
\label{alg:harq}
\begin{algorithmic}[1]
\Require $\hat{z}$, $\hat{y}$, $M^{\mathrm{b}}$, $l$.
\State derive $P_0$, $Q_0$ from Eq. (\ref{eq:PQ}) and $M$ from Eq. (\ref{eq:M_from_Mb}).
\State compute $\tilde{z}^0\coloneqq\hat{\tilde{z}}$, $\hat{\tilde{y}}$ from Eq. (\ref{eq:received_zy}).
\State set $t\leftarrow 0$,\ $P_t\leftarrow P_0$,\ $Q_t\leftarrow Q_0$, $\tilde{z}^0_t\leftarrow\tilde{z}^0$.
\Repeat
    \State $(M_t^{\mathrm{b}})_{ij}\!\leftarrow\!\mathds{1}\{(i,j)\in Q_t\}$;\quad $M_t\!\leftarrow\!M_t^{\mathrm{b}}\otimes\mathbf{1}_{l\times l}$.
    \State $\hat{\tilde{z}}_t \!\leftarrow\! \oplus\!\left\{\mathcal{S}^{-1}\!\big(\mathcal{C}^{-1}(\mathcal{M}^{-1}(\hat z))\big)\big|_{P_t}\ ;\ M_t\right\}$;\quad $\tilde z^{0}_t\!\leftarrow\!\hat{\tilde{z}}_t$.
    \State $d_t \!\leftarrow\! \lvert Q_t\rvert/N$;\quad $s_t\!\leftarrow\! s(\mathrm{SNR},d_t)$;\quad $S_t\!\leftarrow\!\lceil s_t T\rceil$.
    \For{$\gamma \in \{w,0\}$}
        \State $\hat{x}^{(\gamma)}_t\leftarrow\Gamma(\tilde{z}^0_t,M_t;\ \gamma,\ s_t,\ S_t)$;
        \State $\bar{y}^{(\gamma)}_t\leftarrow\mathcal{F}(\hat{x}^{(\gamma)}_t)$;\quad $r^{(\gamma)}_t\leftarrow\Phi\bigl(\bar{y}^{(\gamma)}_t,\hat{\tilde{y}}\bigr)$;
    \EndFor
    \State select \(r_t=\max\{r^{(w)}_t,r^{(0)}_t\}\) and the corresponding \(\hat{x}_t\).
    \If{$r_t<\tau$ \textbf{and} $t<t_{\max}$}
        \State choose $\Delta_t \subseteq Q_t$ uniformly with $\lvert\Delta_t\rvert=\rho N$ and request $\Delta_t$.
        \State $P_{t+1}\leftarrow P_t \cup \Delta_t$;\ $Q_{t+1}\leftarrow Q_t \setminus \Delta_t$;\ $t\leftarrow t+1$.
    \EndIf
\Until{$r_t \ge \tau$ \textbf{or} $t=t_{\max}$}
\State \textbf{return} $\hat{x}_t$
\end{algorithmic}
\end{algorithm}

\subsection{Adaptive Retransmission with Receiver-side Semantic Criterion}

Deciding in advance how many latent blocks to send is difficult, because the useful evidence depends on image content and channel conditions. We adopt a receiver-driven semantic retransmission mechanism in which both the trigger and the stopping rule are defined in the semantic space to match the communication objective. Since the original image is unavailable at the receiver, we use ROUGE-L \cite{lin2004rouge} (denoted by $\Phi$) to score the reconstruction by generating a caption $\bar{y}_t=\mathcal{F}(\hat{x}_t)$ and computing
$r_t=\Phi(\bar{y}_t,\hat{\tilde{y}})\in[0,1]$.
Larger $r_t$ indicates stronger semantic alignment between the reconstruction and the received intent. If $r_t$ meets or exceeds the threshold $\tau$, the session terminates. Otherwise the receiver requests a small number of withheld latent blocks uniformly at random, merges the newly decoded blocks into the currently known latent, and updates the mask using the indicator function $\mathds{1}(\cdot)$ to define the next inpainting region.

Let $P_0$ and $Q_0$ be the initial transmitted and withheld index sets that are obtained from $M^{\mathrm{b}}$, with round index $0$ as their lower subscript and $\lvert P_0\rvert+\lvert Q_0\rvert=N$. At round $t$, the receiver requests a subset $\Delta_t\subseteq Q_{t}$ with $\lvert\Delta_t\rvert=\rho N$, where $\rho$ is the per-round budget ratio. The sets update as
\begin{equation}
\begin{aligned}
Q_{t+1} = Q_{t}\setminus \Delta_t, \qquad
P_{t+1} = P_{t}\cup \Delta_t,
\end{aligned}
\end{equation}
where $t\in\mathbb{N}$ and $t\leq t_{max}$. Only newly requested blocks are transmitted. After decoding, the receiver inserts the newly recovered blocks into the full latent to form $\hat{\tilde z}_t$, with sites in $Q_{t+1}$ kept masked (filled with zeros) for inpainting. In each round, the decoder reruns with the calculated strength $s_t$. To mitigate unstable guidance under noise, the receiver reconstructs two candidates per round, one with guidance scale $w$ and one with $\gamma=0$, and selects the candidate that yields the larger semantic score $r_t$. Algorithm~\ref{alg:harq} summarizes the receiver-side semantic-driven adaptive retransmission. We do not model packet erasures and assume a reliable feedback channel; incorporating link-layer losses is left to future work.


\section{Experimental Results and Discussion}


\subsection{Simulation Settings and Benchmarks}

We use Stable Diffusion v2.1 as the generative backbone $\Gamma$ and BLIP-2 as the receiver-side captioner $\mathcal{F}$ (Algorithm~\ref{alg:harq}). Experiments are conducted on the Flickr30k dataset. Each image is resized and encoded by the Stable Diffusion v2.1 VAE encoder to obtain a latent grid, which is then partitioned into blocks for transmission. The per-round budget is set so that the transmitted latent payload is capped by the protocol parameters in Table~\ref{table:settings}.

To comprehensively evaluate system performance, beyond using ROUGE-L as the receiver-side semantic criterion, we adopt complementary metrics that capture different facets of reconstruction quality. Structural similarity index measure (SSIM) assesses pixel-level fidelity (higher is better). Learned perceptual image patch similarity (LPIPS)
measures perceptual dissimilarity as a distance in deep feature space (lower is better). Fréchet inception distance (FID) quantifies distributional realism via the Fréchet distance between Gaussians fitted to Inception-V3 feature embeddings of generated and reference image sets (lower is better). Finally, CLIP-based image-text similarity (denoted as CLIP-IT) measures image-text semantic alignment as the cosine similarity between CLIP image and text embeddings (higher is better). 
We also report the semantic failure probability $P_{\text{fail}}$, defined as the fraction of test images whose returned reconstruction has an evaluation score below $\tau$ at the maximum allowed round, i.e., $P_{\text{fail}}=\Pr\{r^{\text{eval}}_{t^*}<\tau,\ t^*=t_{\max}\}$.
For evaluation, we use the clean caption $\tilde{y}=\mathcal{F}(x)$ as the reference across all schemes and define
$r^{\text{eval}}_t \triangleq \Phi(\bar{y}_t,\tilde{y})$.
Only the online control loop uses the received caption $\hat{\tilde{y}}$ and its corresponding control score
$r^{\text{rx}}_t \triangleq \Phi(\bar{y}_t,\hat{\tilde{y}})$ for early stopping. We use several baselines:

\begin{itemize}
\item \textbf{No-guidance:} $\mathcal{R}$ does not use the caption $y$ and reconstructs without text conditioning, but runs the same number of rounds as the main scheme for each test image to enable paired-budget comparison.
\item \textbf{Full-mask:} the transmitter sends only the caption $y$ and no latent blocks, i.e., $M^{\mathrm{b}}=\mathbf{1}$ and $t=0$.
\item \textbf{Fixed-budget:} the transmitter sends $N_{\mathrm{fix}}$ latent blocks in a single shot without feedback, where $N_{\mathrm{fix}}$ matches the main scheme's average number of transmitted blocks under the same $(\tau,\rho,t_{\max})$ at each $(\mathrm{SNR},l)$.
\item \textbf{Fixed-round:} the receiver follows the same per-round request rule as the main scheme but always runs $t_{\max}$ rounds without early stopping, serving as a high-overhead reference that transmits substantially more latent and is conceptually related to caption-plus-latent schemes that convey more latent-block evidence \cite{cicchetti2024language}.
\end{itemize}

The inpainting strength $s$ in Algorithm~\ref{alg:harq} is set by the heuristic in Table~\ref{table:settings}. For a fair comparison, the no-guidance and full-mask baselines use the same $s$ as the main scheme under the same $(\mathrm{SNR},d)$. We use the fixed-round scheme (always running $t_{\max}$ rounds) as a budget-constrained upper reference because it removes early stopping while keeping the same per-round request rule. We intentionally exclude the no-mask scheme that transmits all latent blocks, since it is an unconstrained upper bound rather than a rate-comparable baseline.

Here we clarify rate accounting. The transmitted caption is typically about $60$ to $100$ characters in our setting, corresponding to roughly $960$ to $1600$ coded bits after ASCII mapping and rate-$1/2$ channel coding. By contrast, one round request ($\rho N$ latent blocks) carries $131072$ bits under our latent serialization, so the caption contributes only a small fraction of the total transmitted bits and does not materially bias comparisons with the no-guidance scheme. Note that this is on top of compressing the image into latent space; the adaptive scheme still transmits substantially less than sending the full latent grid.

\begin{table}[t]
\centering
\caption{Key simulation settings.}
\begin{tabular}{l|l}
\hline
Image size & $3\times512\times512$ \\
Latent size & $4\times64\times64$ \\
Block size on latent grid & $l\in\{4,8\}$ \\
Initial sending rate & $\kappa_0=0.125$ \\
Per-round budget & $\rho=0.0625$ \\
Channel model & AWGN channel; SNR $\in[5,10]$ dB \\
Guidance scale & $w=9$ (fixed) \\
Inpainting strength & $s=\min\{1,\sqrt{d}(1\!-\!0.1\!\tanh(\mathrm{SNR}\!-\!10))\}$ \\
DDIM maximum steps & $T=50$ \\
Max rounds & $t_{\max}=6$ \\
\hline
\end{tabular}
\label{table:settings}
\end{table}

\subsection{Experiment Results}

\begin{figure*}
    \centering
    \subfigure[]{\includegraphics[height=2.71cm]{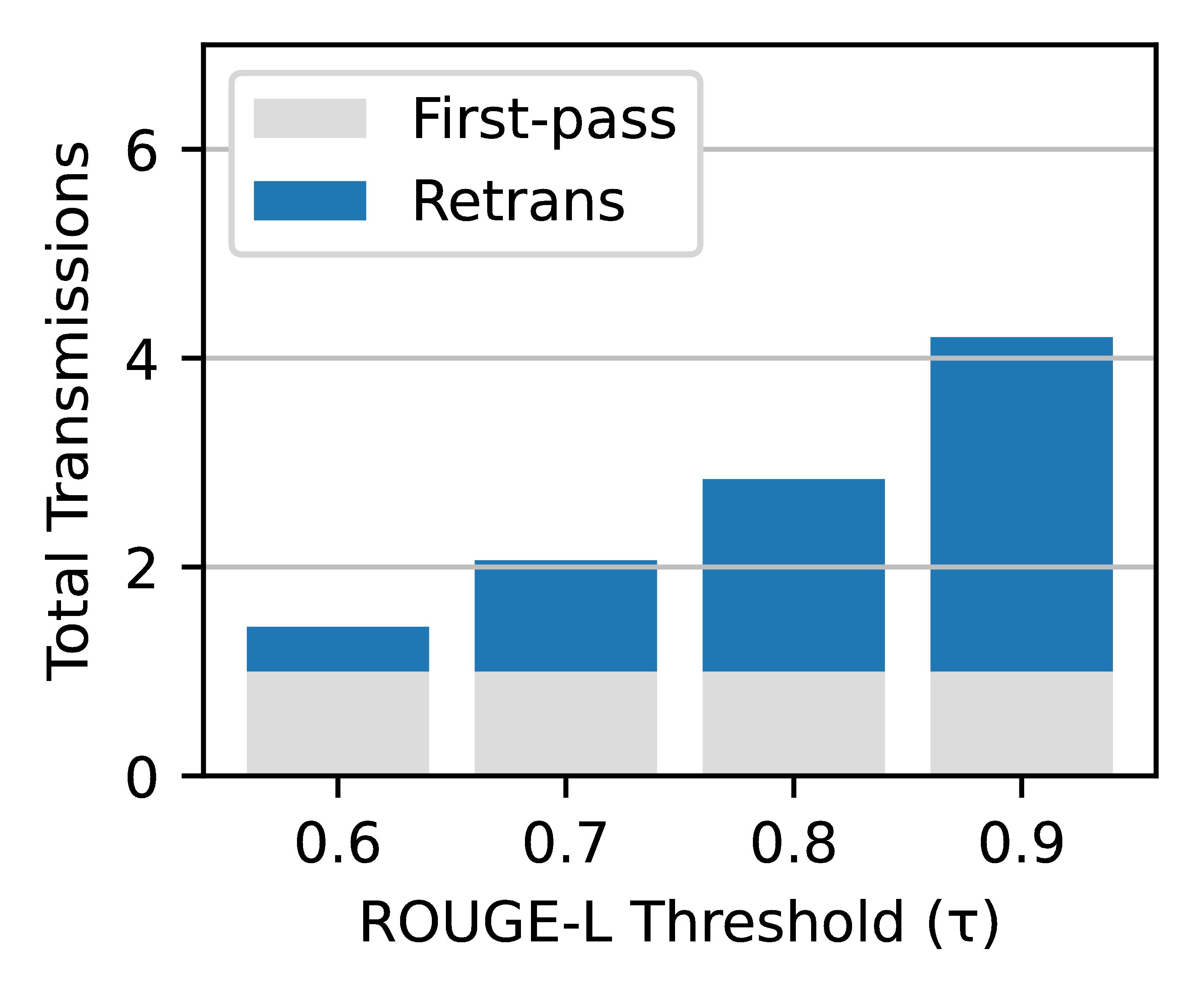}}\hspace{1em}
    \subfigure[]{\includegraphics[height=2.71cm]{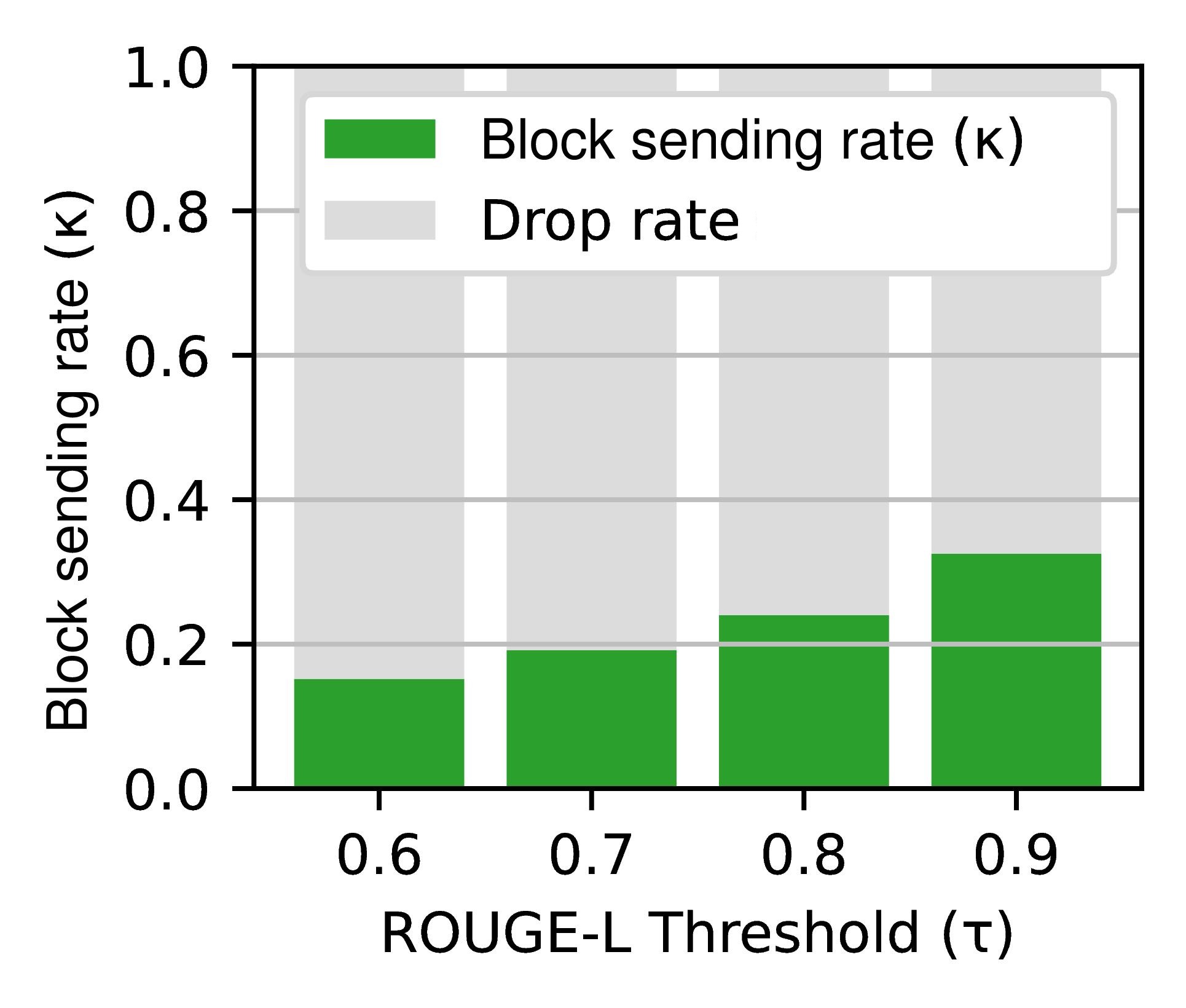}}\hspace{1em}
    \subfigure[]{\includegraphics[height=2.71cm]{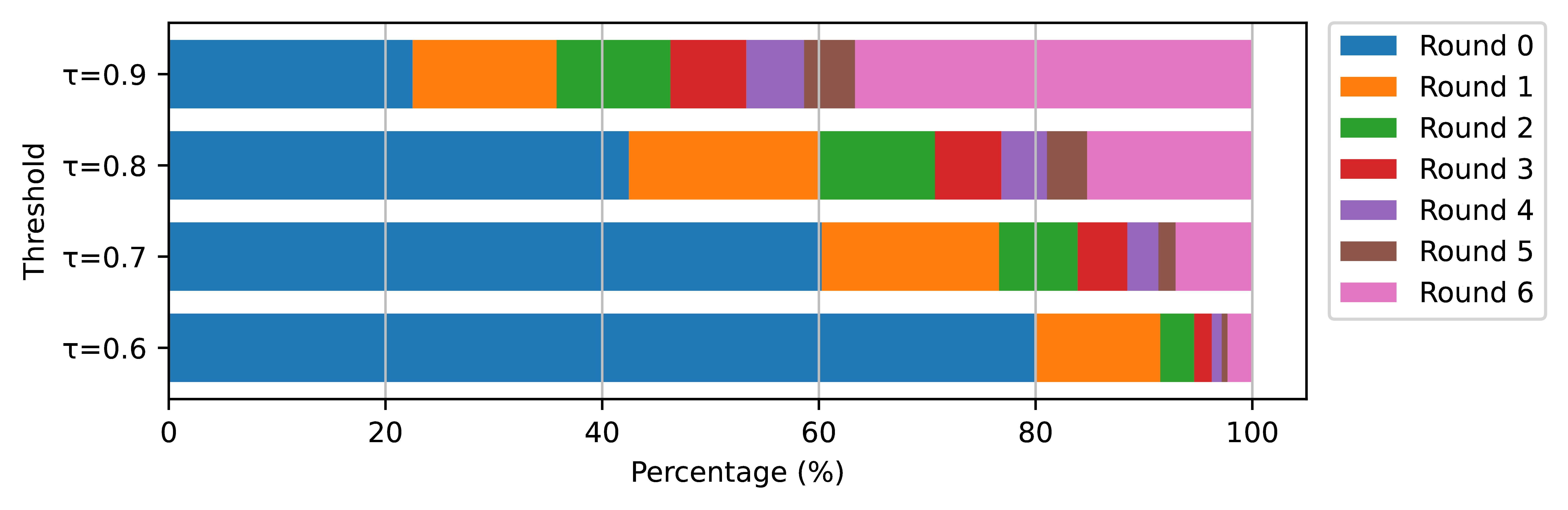}}
    \caption{Effect of the ROUGE-L threshold $\tau$. (a) Total number of transmission rounds (including the first pass and feedback rounds). (b) Final block sending rate $\kappa$. (c) Histogram of termination round index $t$.}
    \label{fig:fig-bars}
\end{figure*}

\begin{figure*}
    \centering
    \subfigure[ROUGE-L]{\includegraphics[height=3.88cm]{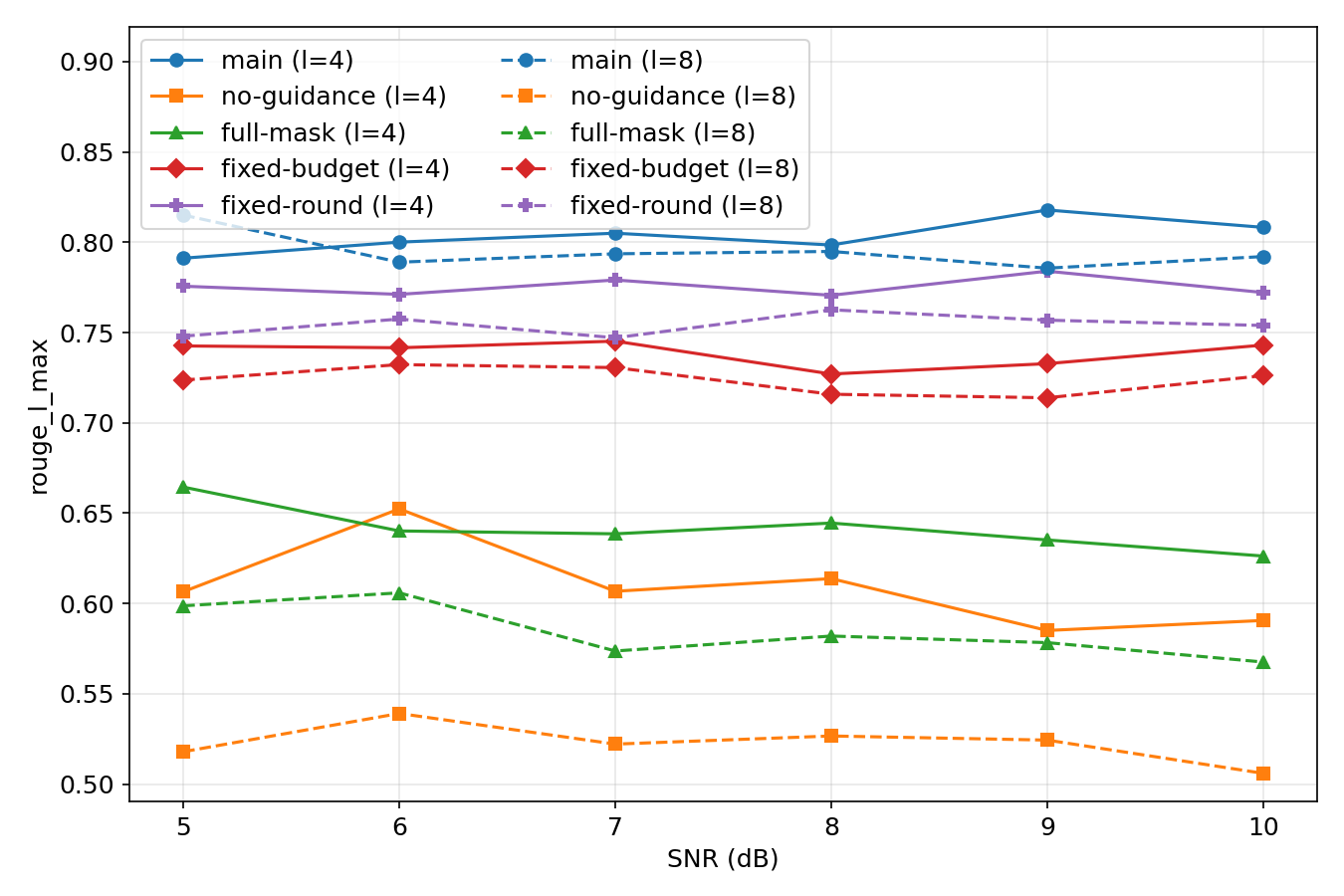}}
    \subfigure[CLIP-IT]{\includegraphics[height=3.88cm]{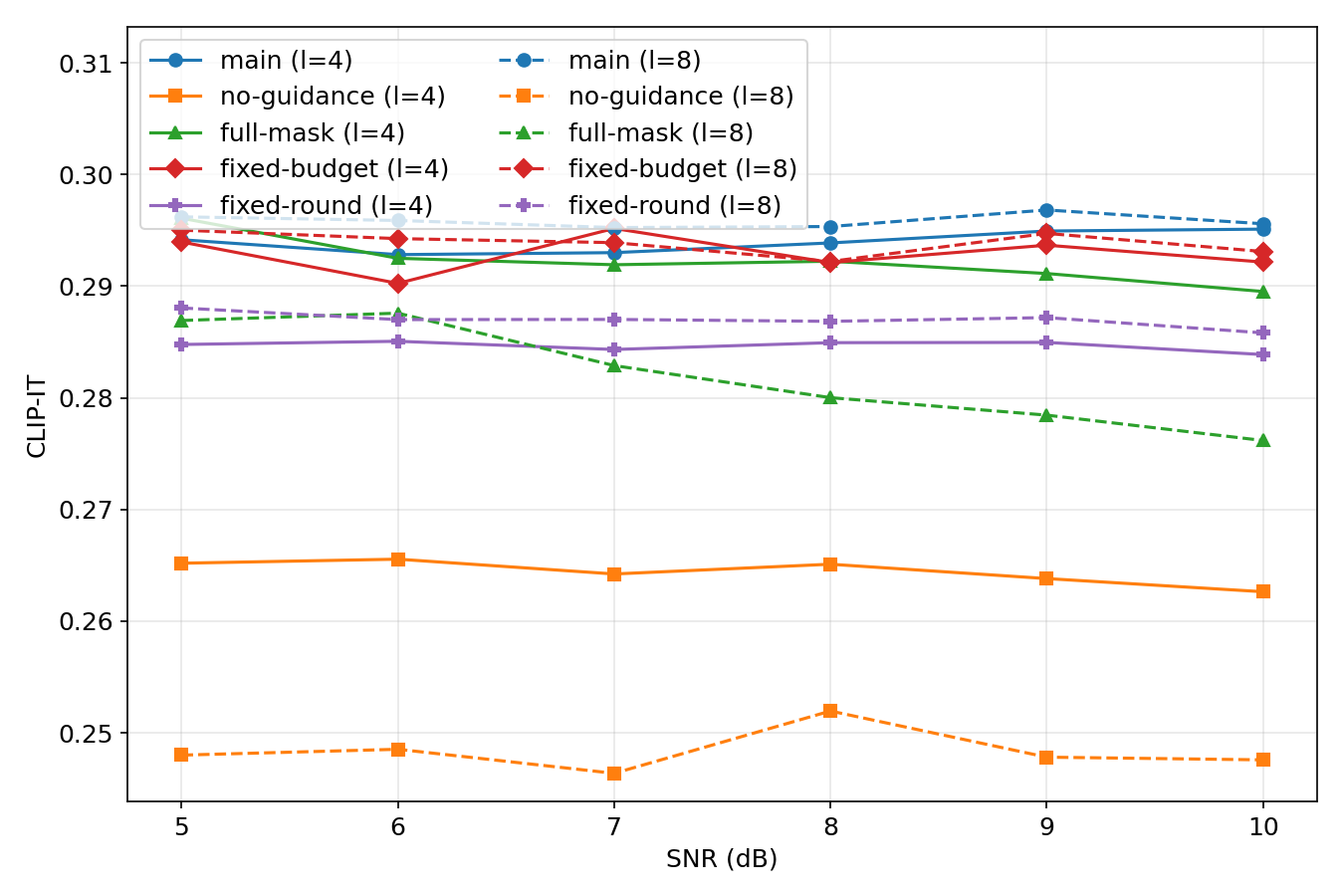}}
    \subfigure[SSIM]{\includegraphics[height=3.88cm]{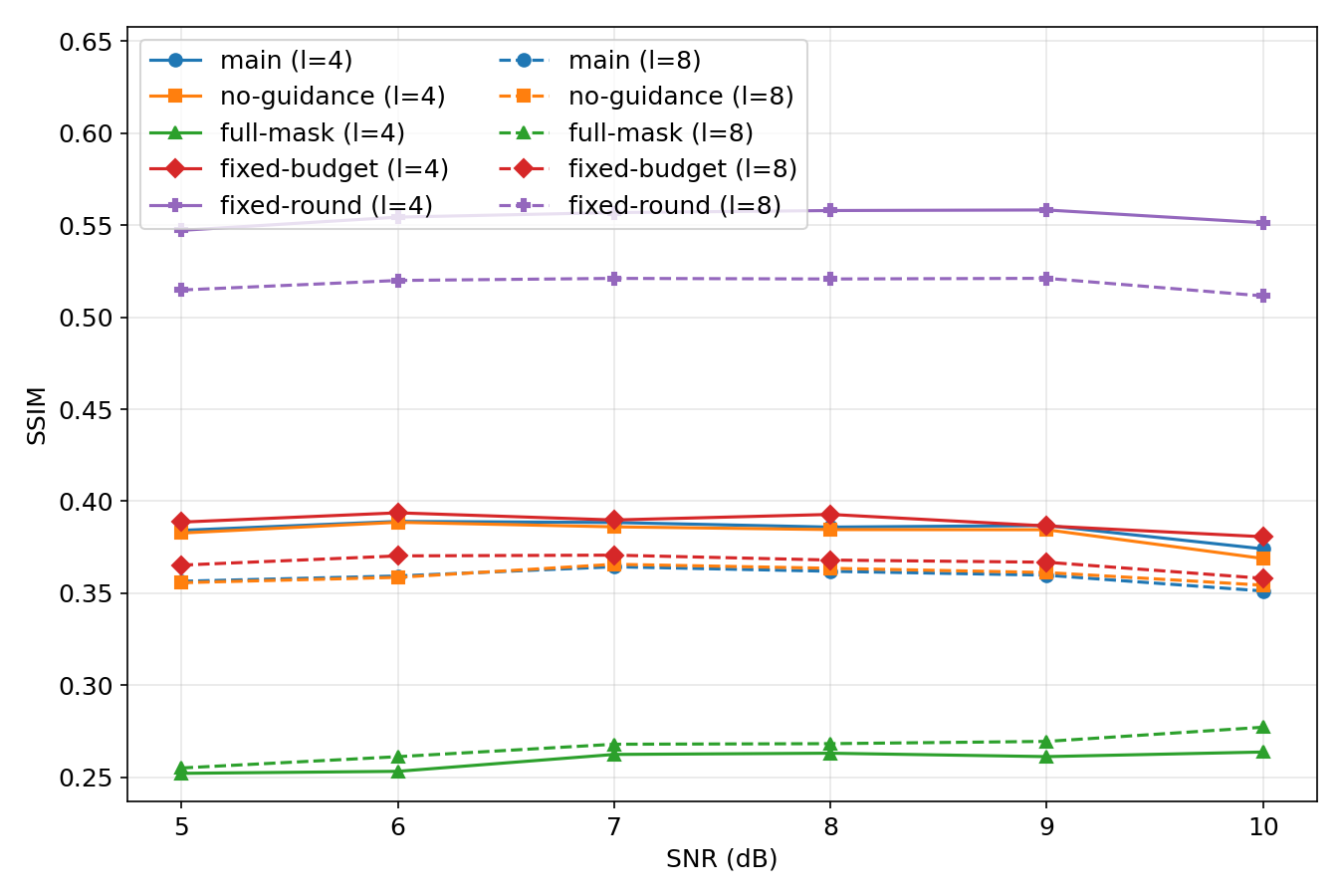}}\\
    \subfigure[LPIPS]{\includegraphics[height=3.88cm]{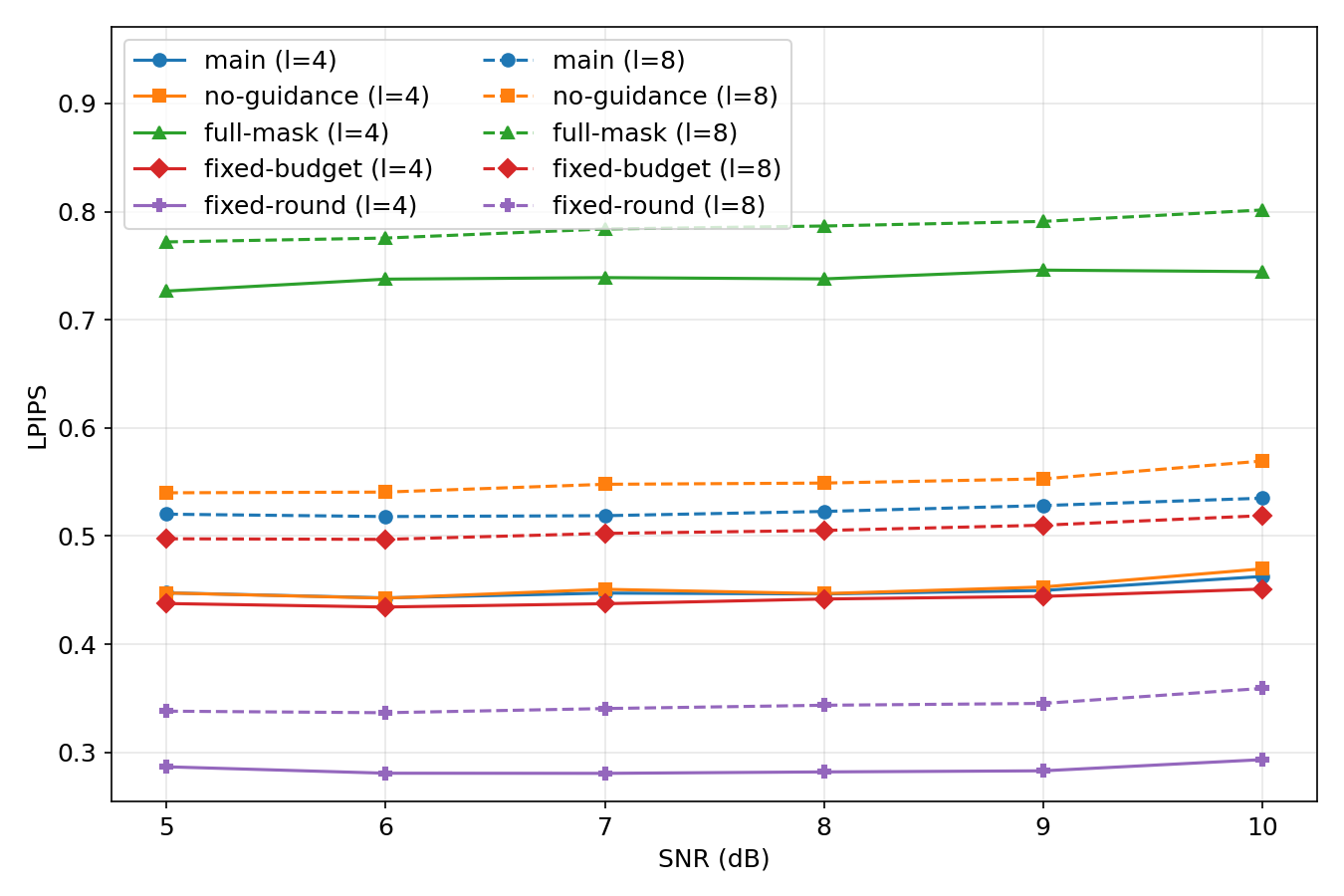}}
    \subfigure[FID over SNR]{\includegraphics[height=3.88cm]{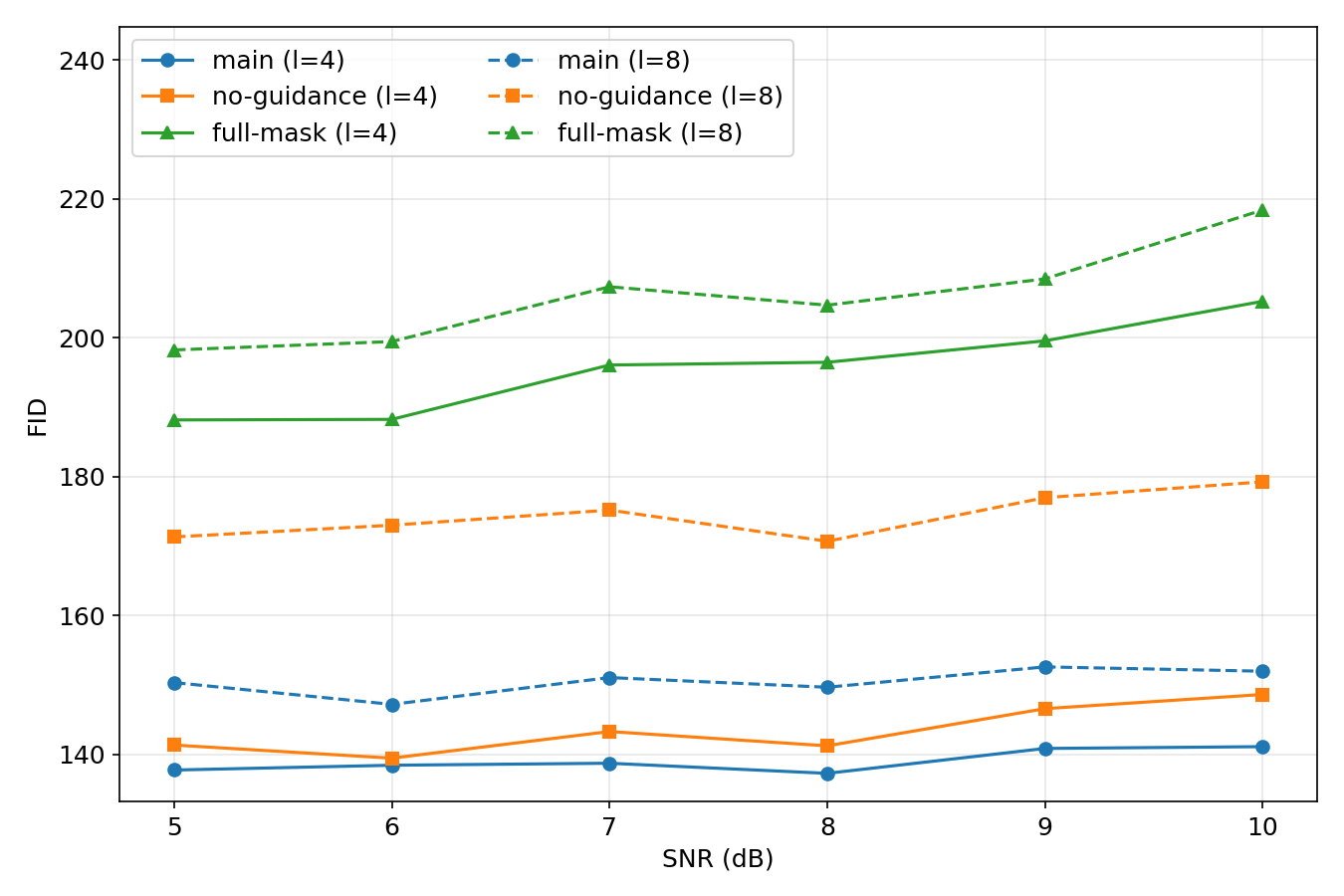}}
    \subfigure[FID over $\tau$]{\includegraphics[height=3.88cm]{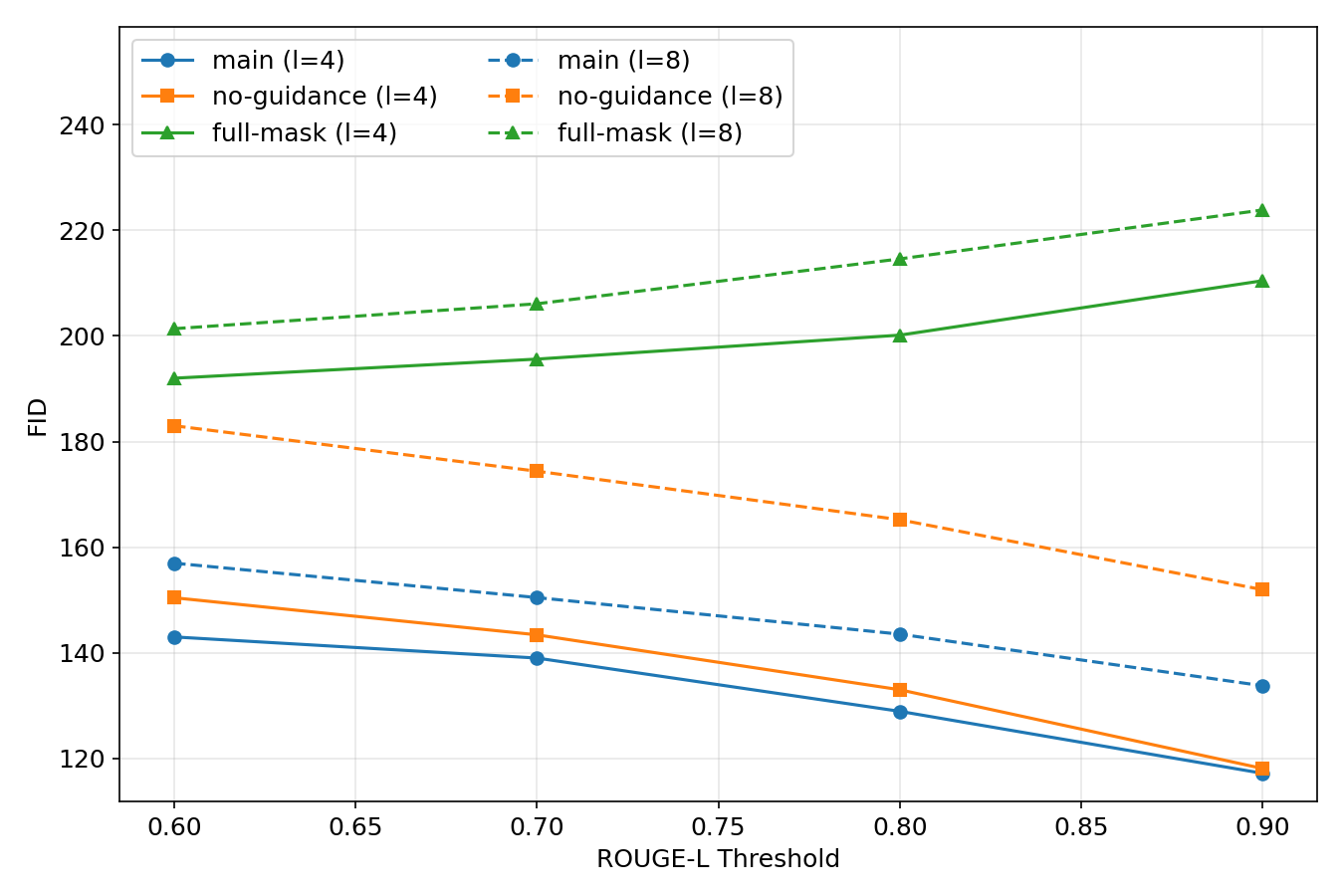}}
    \caption{System performance. (a)(b)(c)(d) metrics over SNR values at $\tau=0.7$. (e)(f) FID performance over SNR and $\tau$ values.}
    \label{fig:fig-metrics}
\end{figure*}

Fig.~\ref{fig:fig-bars} shows that the threshold $\tau$ directly controls how conservative the closed loop is in early stopping. As $\tau$ increases, the receiver requires stronger semantic agreement before stopping, so the termination round index shifts to later rounds and the average number of rounds increases. This increases the total number of requested latent blocks and raises the sending ratio $\kappa$, i.e., more semantic content is ultimately delivered. The histogram further indicates that most sessions terminate within a few rounds for all tested $\tau$, while larger $\tau$ mainly increases the probability of late termination.

Fig.~\ref{fig:fig-metrics}(a)-(d) report the average performance at $\tau=0.7$ for two block sizes $l\in\{4,8\}$, comparing the main scheme with four baselines. On semantic-oriented metrics, the main scheme achieves the highest ROUGE-L and is among the top CLIP-IT performers across SNR values, with fixed-budget being close on CLIP-IT. The no-guidance baseline (latent-only) degrades substantially, indicating that without a textual constraint the generative prior can drift in high-level content even when partial latent evidence is available. The full-mask (caption-only) baseline lacks source-specific latent evidence, leading to the weakest pixel/perceptual quality and poorer semantic alignment among the baselines. Notably, while fixed-round scheme runs more rounds, it does not necessarily increase the caption-consistency score. Each round re-runs stochastic diffusion inpainting with additional revealed blocks, and the sampler is not designed to monotonically improve a language-based criterion, so the caption-consistency score is not guaranteed to improve monotonically, since diffusion-based inpainting is stochastic and can yield different plausible completions even when more evidence is revealed \cite{lugmayr2022repaint}.

Comparing the main scheme with the fixed-budget one-shot baseline isolates the benefit of receiver-driven adaptation under the same average latent payload. Although the fixed-budget baseline transmits $N_{\mathrm{fix}}$ blocks matched to the main scheme on average, it cannot adapt its payload to sample difficulty and therefore yields lower ROUGE-L. This is the key advantage of the closed loop. It spends extra blocks only when the receiver-side semantic criterion is not met, while stopping early on easier samples.

On pixel/perceptual fidelity (SSIM/LPIPS), the fixed-round scheme performs best because it always runs $t_{\max}$ rounds and thus accumulates substantially more retransmitted latent blocks than the early-stopping schemes (as shown in Fig.~\ref{fig:fig-bars}(a), where the main scheme typically terminates after far fewer rounds). This curve should therefore be interpreted as a high-overhead reference rather than a rate-comparable baseline. In contrast, the main and fixed-budget schemes are close on SSIM/LPIPS, indicating that matching the average latent payload largely closes the pixel/perceptual gap; the main scheme primarily differs by reallocating the same budget across images via semantic-driven stopping.

Fig.~\ref{fig:fig-metrics}(e) shows that FID mildly increases with SNR for some configurations. This is consistent with our heuristic that decreases the inpainting strength and therefore the number of denoising steps at higher SNR. The sampler relies more on the received latent evidence and performs fewer corrective diffusion updates, which can slightly worsen dataset-level feature statistics. Fig.~\ref{fig:fig-metrics}(f) further shows that increasing $\tau$ improves FID for the main and no-guidance schemes because more evidence is delivered before stopping, while the caption-only full-mask baseline degrades as $\tau$ increases.

\begin{figure}[!t]
    \centering
    \includegraphics[height=3.88cm]{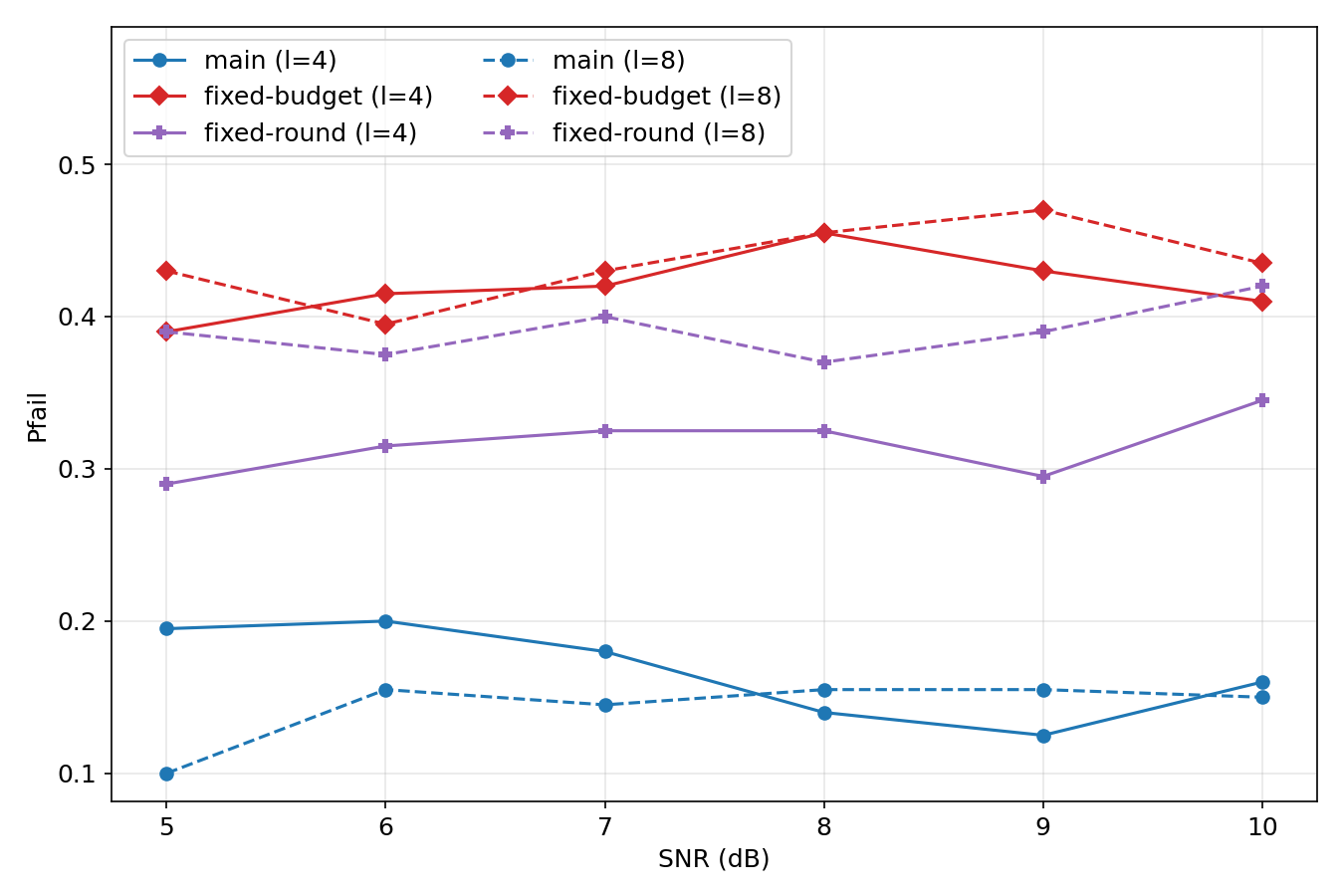}
    \caption{Semantic failure probability $P_{\text{fail}}$ versus SNR at $\tau=0.7$.}
    \label{fig:pfail}
\end{figure}

Fig.~\ref{fig:pfail} reports $P_{\text{fail}}$ versus SNR at $\tau=0.7$. The main scheme consistently achieves the lowest failure probability, since feedback injects additional latent evidence only when the semantic score is insufficient. Fixed-round can show higher $P_{\text{fail}}$ despite sending more blocks, because the caption-consistency score is not guaranteed to improve monotonically under stochastic diffusion re-inpainting \cite{lugmayr2022repaint}.

\begin{figure}[t]
    \centering
    \setlength{\tabcolsep}{3pt}
    \renewcommand{\arraystretch}{1.0}
    \footnotesize
    \begin{tabular}{@{} r m{1.6cm} @{\hspace{4pt}} r m{1.6cm} @{\hspace{4pt}} r m{1.6cm} @{}}
        (a) & \includegraphics[height=1.6cm]{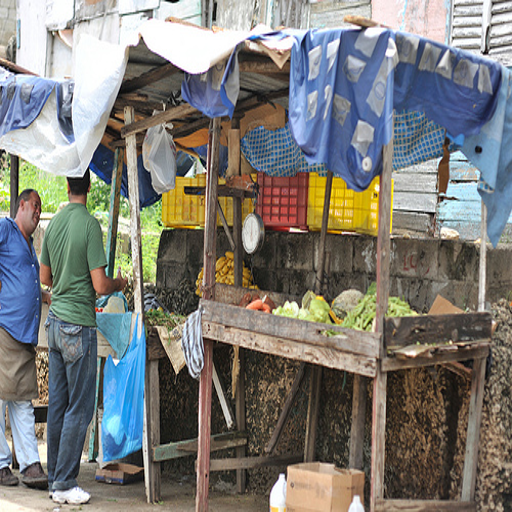} &
        (b) & \includegraphics[height=1.6cm]{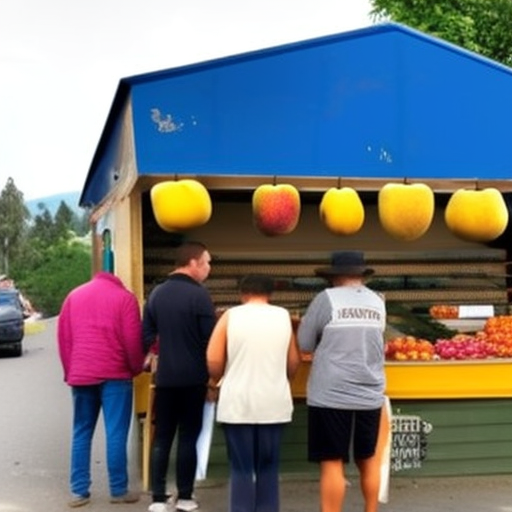} &
        (c) & \includegraphics[height=1.6cm]{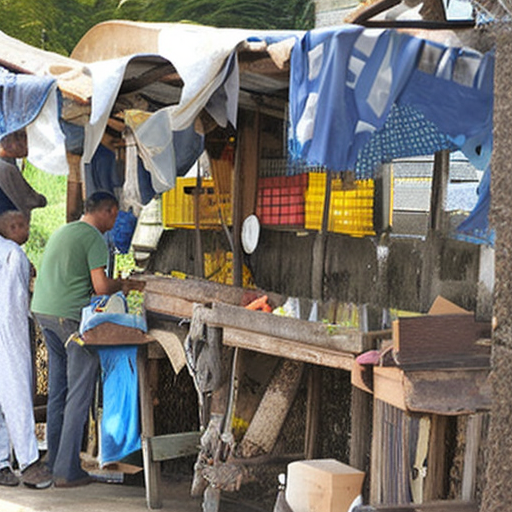} \\
        (d) & \includegraphics[height=1.6cm]{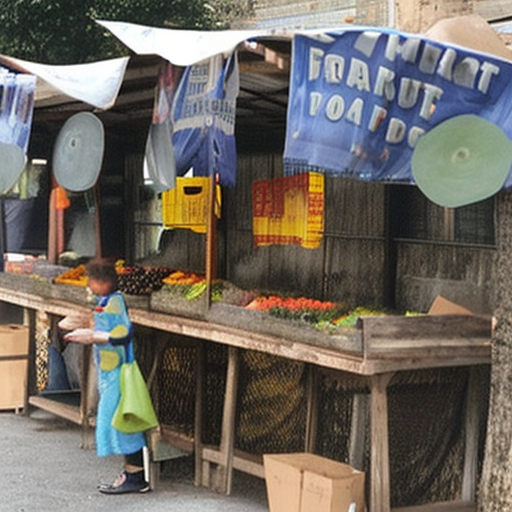} &
        (e) & \includegraphics[height=1.6cm]{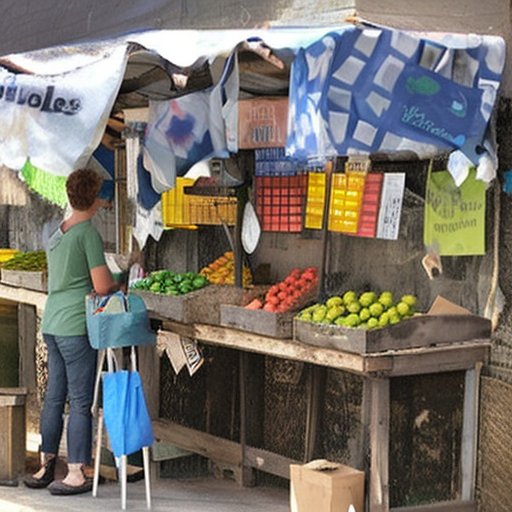} &
        (f) & \includegraphics[height=1.6cm]{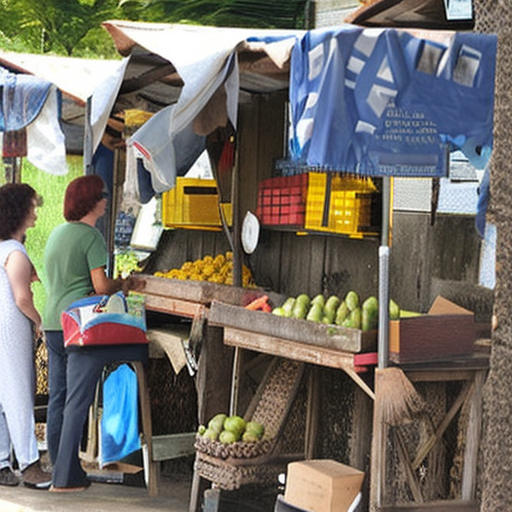} \\
    \end{tabular}
    \caption{Qualitative example at $\kappa=0.25$ (SNR $=7$ dB, $l=4$, $\tau=0.7$): (a) original, (b) full-mask, (c) no-guidance at $t=2$, (d)-(f) main at $t=0,1,2$.}
    \label{fig:compare}
\end{figure}

Fig.~\ref{fig:compare} shows a representative example at $\kappa=0.25$. The full-mask baseline produces a plausible image but lacks source-specific details, while the no-guidance baseline tends to preserve local textures yet may drift in high-level semantics. In contrast, the main scheme progressively refines the reconstruction across rounds as additional latent blocks are requested, and the reconstruction becomes increasingly consistent across rounds, reducing inconsistent regions while keeping the content aligned with the caption constraint.

Finally, we comment on complexity. Runtime is dominated by diffusion sampling, and the overall cost scales with the cumulative number of denoising steps across rounds, i.e., $\sum_t S_t$. VAE encode/decode, caption generation, and score computation run once per round and add a small constant overhead relative to sampling. Memory is dominated by the fixed model parameters and U-Net activations during sampling, and thus remains essentially constant across retransmission rounds.


\section{Conclusion}

This paper proposes a receiver-driven semantic retransmission framework for image transmission with a generative receiver. The transmitter sends a short caption and an initial sparse subset of VAE latent blocks, while the receiver performs caption-conditioned latent diffusion inpainting and iteratively requests additional blocks until a receiver-side semantic criterion is met or a round limit is reached.
Experiments on Flickr30k over AWGN channels confirm a tunable rate-quality tradeoff. Compared with budget-matched one-shot sparse transmission, closing the loop yields higher ROUGE-L semantic consistency and fewer semantic failures, and it generally requires fewer latent blocks than an always-on retransmission scheme.
Overall, the results indicate that receiver-side semantic control can improve reliability for generative receivers under a fixed average evidence budget. Future work will consider importance-aware block selection, unreliable captions/feedback, and extensions to video and multi-user settings.

\bibliographystyle{IEEEtran}
\bibliography{refs}

\end{document}